\documentstyle[fleqn]{article}
\newcommand{\bg}{\begin{equation}}
\newcommand{\eg}{\end{equation}}

\def\diag{\mbox{diag }}

\def\id{\mbox{\rm 1\hskip-.25em l}}

\def\LD{{\cal L}}

\def\lplus{\supset\mkern -19mu\hbox{\small +}\mkern 10mu}

\newcommand{\mb}[1]{\mbox{\boldmath $#1$}}
\newcommand{\mr}[1]{\mbox{\bf #1}}
\begin{document}
\centerline{\normalsize\bf
LEPTONIC GENERATION MIXING, NONCOMMUTATIVE GEOMETRY}
\baselineskip=16pt
\centerline{\normalsize\bf AND SOLAR NEUTRINO FLUXES}
\vspace*{2cm}
\centerline{\footnotesize
R. H\H au\ss ling$^\dagger$, M. Paschke$^\ddagger$,
and F. Scheck$^\ddagger$}
\baselineskip=13pt
\centerline{\footnotesize\it
{}$^\dagger$ Institut f\H ur Theoretische Physik, Universit\H at Leipzig,
D-04109 Leipzig}
\centerline{\footnotesize\tt
Haeussli@tph204.Physik.Uni-Leipzig.DE}
\baselineskip=12pt
\centerline{\footnotesize\it
{}$^\ddagger$ Institut f\H ur Physik - Theoretische Elementarteilchenphysik}
\centerline{\footnotesize\it
Johannes Gutenberg-Universit\H at, D-55099 Mainz}
\centerline{\footnotesize\tt
(name)@dipmza.Physik.Uni-Mainz.DE}
\normalsize\baselineskip=15pt
\vspace*{3cm}
\begin{abstract}
Triangular mass matrices for neutrinos and their charged partners contain
full information on neutrino mixing in a most concise form. Although
the scheme is general and model independent, triangular matrices are
typical for reducible but indecomposable representations of graded Lie
algebras which, in turn, are characteristic for the standard model in
noncommutative geometry. The mixing matrix responsible for neutrino 
oscillations is worked out analytically for two and three lepton families.
The example of two
families fixes the mixing angle to just about what is required by the
Mikheyev-Smirnov-Wolfenstein resonance oscillation of solar neutrinos.
In the case of three families we classify all physically plausible 
choices for the neutrino mass matrix and derive interesting bounds on
some of the moduli of the mixing matrix.
\end{abstract}
\vfill
\newpage
\noindent
{\bf 1.} While for many years neither direct searches
nor terrestrial oscillation
experiments have given any positive evidence for nonvanishing neutrino
masses, the intense experimental investigation of solar neutrino fluxes
significantly changed this situation. Indeed, the four solar neutrino
experiments which test different parts of the energy spectrum and
all of which find fluxes reduced as compared to the ones
expected on the basis of standard solar model calculations \cite{Bahcall}
and non-oscillating electron neutrinos, seem to lend support to the
hypothesis of resonance enhanced oscillation \cite{MikSmir,Wolf} within the
sun's interior. The flux observed by Davies et al. using inverse
$\beta$-decay on ${}^{37}$Cl which is about one third of the expected flux
and which is sensitive to the high-energy neutrinos from boron, the flux
seen in the two experiments with ${}^{71}$Ga, {\sc gallex} and  {\sc sage},
which is about 60 percent of the expected, and the {\sc kamiokande} results
on $\nu_e$-electron scattering, seem compatible with the
Mikeyev-Smirnov-Wolfenstein mechanism calculated from standard solar
parameters and two-generation mixing with squared mass difference
and mixing angle (vacuum values) in the range \cite{Brussel}
\bg \label{m2th}
\Delta m^2\approx 6\cdot 10^{-6} \mbox{eV}^2\, ,\qquad
\sin^2\theta\approx 0.006\, .
\eg

In the meantime further experimental hints at nonvanishing neutrino
masses is provided by observation of atmospheric neutrinos \cite{atmn},
and by an accelerator experiment at LAMPF \cite{lanl}. We note that
empirical analyses of all five experiments indicate that they cannot
all be fit with three families of leptons \cite{BGG}.

Conventional gauge theories of electroweak interactions are based on
structure groups which are compact Lie groups. Fermions are classified
in finite-dimensional, unitary (in general reducible) representations
$\rho_l$
of the structure group. As such representations are always fully reducible
the gauge bosons of electroweak interactions, a priori, can mediate only
between members of the same, given fermionic family. Any transition
between states pertaining to two different families must be due to
a new element in the theory.

We adopt the usual attitude in assuming that the
mixing of fermionic states with equal charges but different flavours is
solely due to the mismatch between weak interaction states and mass
eigenstates. This means that the mixing matrix relevant for the 
charged current (CC) interactions of the standard model is calculable from
the mass matrices. Making use of the fact that CC interactions involve
only left-chiral fields we recently showed that the full information
contained in any nonsingular mass matrix can be described
by means of a (lower) triangular matrix~\cite{HAS}. The triangular form
which is obtained by (unobservable) unitary transformations of right-chiral
fields, is the most economic and concise description of family
mixing. As shown in ref.~\cite{HAS} it corresponds effectively to
classifying the lepton families in reducible but indecomposable
representations of the type
\bg \label{semi}
\rho_l\lplus\rho_l \qquad\mbox{or}
\qquad\rho_l\lplus \rho_l\lplus\rho_l \, ,
\eg
for two and three families, respectively, where the ``semi-sum'' means
that the right-hand representation space is an invariant subspace while
the left one(s) is (are) not.

Although our analysis is completely general, we note that
representations of the type (\ref{semi}) are not unexpected. Indeed,
extensions of gauge theories as obtained in the framework of noncommutative
geometry add more structure to them and, in particular, constrain the
fermionic sector in a novel and physically relevant way.
In the approach proposed by Connes and Lott \cite{CoLo}
it is postulated that the Dirac-Yukawa operator contain the physical quark
and lepton mass matrices, including the quark mixing pattern. The bosonic
sector as derived from that operator exhibits spontaneous symmetry breaking
with the distinctive feature that the Higgs potential would vanish if
generations did not mix or were degenerate.

The construction proposed by the Mainz-Marseille group \cite{CES,MaiMar},
on the other hand, starts from the
underlying graded differential algebra and yields the bosonic sector,
including spontaneous symmetry breaking, without resorting to the fermion
content of the theory. The Dirac-Yukawa operator is a derived object whose
detailed form, unlike the Connes-Lott construction, is not an input.
The graded gauge potential of this model comprises the gauge bosons
as well as the Higgs field and is seen to fall into the adjoint
representation of the graded Lie algebra $su(2\,\vert\, 1)$, viz.
\begin{eqnarray*}
[Y_0=0,I_0=1] & \longrightarrow & (I=1,Y=0)\oplus (I=0,Y=0)\oplus\\
 & & (I=\frac{1}{2},Y=1)\oplus (I=\frac{1}{2},Y=-1)\, .
\end{eqnarray*}

The right-hand side gives its decomposition with respect to
$Lie\left( SU(2)\times U(1)\right)$ and describes the four gauge bosons,
the Higgs doublet and its antidoublet, respectively.
As explained in earlier publications \cite{MaiMar,HPS} it is natural to use
this algebra as
a classifying algebra for quarks and leptons as well, its action on the
fields of the standard model being understood in the sense of what we called
a {\em weak\/} symmetry in \cite{HPS}. A most remarkable feature of
$su(2\,\vert\, 1)$ is the existence of reducible but indecomposable
representations which are semi-direct sums of two or three copies of the
fundamental representations that correctly describe the $SU(2)\times U(1)$
quantum numbers of quarks and leptons, respectively \cite{CES}. Such
representations, in generation space, have a block triangular form and,
therefore, provide a natural place for generational state mixing.
In \cite{HS} we explored this possibility for quark mass matrices and
CKM mixing. Although based on very different assumptions we showed that
this scheme leads to the mixing pattern favoured by
phenomenology, with the
bonus of providing analytical formulae for the mixing matrix elements in
terms of quark masses and a small number of free parameters.

In this letter we study the case of nontrivial neutrino mass matrices,
in the general framework of triangular mass matrices,
and their consequences for neutrino oscillations including the solar neutrino
fluxes. With a pausible physical assumption on the origin of state
mixing, the case of two families contains no free parameters and
yields a mixing pattern in fair agreement with the resonance oscillation
needed to explain the solar neutrino puzzle. The case of three families
contains more freedom and can be analyzed in terms of complete sets of
neutrino oscillation experiments. We then illustrate this general
framework by the Mainz-Marseille model and show that our physics
assumptions are quite natural within this model based on noncommutative
geometry.\\[4pt]
{\bf 2.} When written out in terms of individual neutral and charged fields
of definite chirality the mass terms are
\bg \label{mass2}
\LD_{mass}=\frac{1}{2}\sum_{i=1}^3\sum_{k=1}^i\left(
\overline{\nu_L^{(i)}}T_{ik}^{(\nu )}\nu_R^{(k)}+
\overline{l_L^{(i)}}T_{ik}^{(l)}l_R^{(k)}\right)
+\mbox{ h.c.}\, .
\eg
The matrices $T$ are lower triangular matrices given by
\bg  \label{tria}
T^{(\nu )}=\left(\begin{array}{ccc}
        a_{11} & 0 & 0 \\ a_{21} & a_{22} & 0 \\ a_{31} & a_{32} & a_{33}
        \end{array}\right)\quad
T^{(l)}=\left(\begin{array}{ccc}
        d_{11} & 0 & 0 \\ d_{21} & d_{22} & 0 \\ d_{31} & d_{32} & d_{33}
        \end{array}\right) 
\eg
for the neutrinos and the charged leptons, respectively.

Without loss of generality, weak interaction and mass eigenstates can be
chosen such that all entries in the triangle matrices (\ref{tria}) except
$a_{31}$ and $d_{31}$ are real \cite{HS}.
Changing notations slightly we
write the latter as
\[
a_{31}e^{i\Phi^{(\nu )}}\quad\mbox{and}\quad
d_{31}e^{i\Phi^{(l)}}\, ,
\]
so that from now on all $a_{ik}$ and all $d_{ik}$ are real. The
characteristic polynomial for the squared mass matrix
$\left( T^{(\nu )}T^{(\nu )\,\dagger}\right)$ yields the following
relations for the parameters
\bg \label{npol1}
m_1^2m_2^2m_3^2=a_{11}^2a_{22}^2a_{33}^2\, ,
\eg
\begin{eqnarray} 
     m_1^2m_2^2+m_2^2m_3^2+
     m_3^2m_1^2 & = &
     a_{11}^2a_{22}^2+a_{22}^2a_{33}^2+a_{33}^2a_{11}^2+
     \nonumber\\
 & & a_{11}^2a_{32}^2+a_{22}^2a_{31}^2+a_{33}^2a_{21}^2+
     \nonumber\\
\label{npol2}
 & & a_{21}^2a_{32}^2-2a_{22}a_{21}a_{31}a_{32}\cos\Phi^{(\nu )}  \, ,
\end{eqnarray}
\bg \label{npol3}
m_1^2+m_2^2+m_3^2=a_{11}^2+a_{22}^2+a_{33}^2+
                      a_{21}^2+a_{31}^2+a_{32}^2\, .
\eg
 Similarly, the
 characteristic polynomial for
 $\left( T^{(l)}T^{(l)\,\dagger}\right)$ yields
 \bg \label{lpol1}
 m_e^2m_\mu^2m_\tau^2=d_{11}^2d_{22}^2d_{33}^2\, ,
 \eg
 \begin{eqnarray}
 m_e^2m_\mu^2+m_\mu^2m_\tau^2+m_\tau^2m_e^2 & = &
     d_{11}^2d_{22}^2+d_{22}^2d_{33}^2+d_{33}^2d_{11}^2+
     \nonumber\\
 & & d_{11}^2d_{32}^2+d_{22}^2d_{31}^2+d_{33}^2d_{21}^2+\nonumber\\
 \label{lpol2}
 & & d_{21}^2d_{32}^2-2d_{22}d_{21}d_{31}d_{32}\cos\Phi^{(l)}  \, ,
 \end{eqnarray}
 \bg \label{lpol3}
 m_e^2+m_\mu^2+m_\tau^2=d_{11}^2+d_{22}^2+d_{33}^2+
                      d_{21}^2+d_{31}^2+d_{32}^2\, .
 \eg

We now make the following, physical assumption: the mass differences
between different families are due to electroweak interactions only.
This means that without these interactions the masses of the electron,
muon and tauon would be degenerate, as would the masses of $\nu_e$,
$\nu_\mu$ and $\nu_\tau$.
When the interaction is present, its effect {\em within\/}
each generation will be the same, the observed splittings of the masses
will be due to the off-diagonal, inter-family blocks in the
mass terms (\ref{mass2}), (\ref{tria}). 
For neutrinos it seems natural to assume further
that their primordial masses vanish. In terms of the parameters in
eq.~(\ref{tria}) this means 
\bg \label{degen}
a_{11}=a_{22}=a_{33}=0\, ,\quad
d_{11}=d_{22}=d_{33}=\left( m_em_\mu m_\tau\right)^{1/3}\, .
\eg

As an example consider the case of two families for which the matrices
(\ref{tria}) simplify to
\bg  \label{tria2}
T^{(\nu )}=\left(\begin{array}{cc}
        a_{11} & 0 \\ a_{21} & a_{22} 
        \end{array}\right)\quad
T^{(l)}=\left(\begin{array}{cc}
        d_{11} & 0 \\ d_{21} & d_{22} 
        \end{array}\right)\, .
\eg

These assumptions imply the following choice of the parameters \cite{HS}:\\
(a) for the neutrino sector
\bg \label{ch21}
a_{11}=0\quad a_{22}=0\quad a_{21}=\pm m_2\, ,
\eg
which lead to the mass eigenvalues $m_1=0$, $m_2\not= 0$, and\\
(b) for the charged lepton sector
\bg \label{ch22}
d_{11}=d_{22}=\sqrt{m_em_\mu}\quad d_{21}=\pm (m_\mu -m_e)\, .
\eg

The mixing matrix is obtained in close analogy to the case of quark
mixing. Let $V^{(\nu)}$ and $V^{(l)}$ be the unitary matrices
which diagonalize the hermitean products
$T^{(\nu )}T^{(\nu )\,\dagger}$ and $T^{(l)}T^{(l)\,\dagger}$,
respectively, i.e. 
  $V\left( TT^\dagger\right)V^\dagger =\diag (m_1^2,m_2^2,m_3^2)$ 
for $(\nu )$ and $(l)$. 
The mixing matrix that determines possible oscillations
is then
\bg \label{vmix}
V_{mix}=V^{(\nu )}V^{(l)\,\dagger}\, .
\eg
In the case of two families it depends on one real angle only
whose cosine is
\bg \label{th}
\cos\theta =\frac{\sqrt{m_1m_e}+\sqrt{m_2m_\mu}}
                 {\sqrt{(m_1+m_2)(m_e+m_\mu)}}=
            \sqrt{\frac{m_\mu}{m_e+m_\mu}}\, .
\eg
{}From this we obtain the result
\bg \label{sin2}
\sin^2\theta =\frac{m_e}{m_e+m_\mu}=0.0048\, .
\eg

Already this first estimate based on two families of leptons is in fair
agreement with the value (\ref{m2th}) needed for solar resonance
oscillation to occur. The mass eigenvalue of the second neutrino remains
undetermined from our analysis. The empirical result (\ref{m2th}) requires it to
be of the order of $10^{-3}$ eV.\\[4pt]
{\bf 3.} Turning now to the more realistic case of
three families we first discuss
the neutrino sector. Making use of the assumption (\ref{degen})
eq.~(\ref{npol1}) implies that at least one neutrino remains massless,
say, $m_1=0$, while eqs.~(\ref{npol2}) - (\ref{npol3}) simplify to
\bg \label{neu}
m_2^2m_3^2=a_{21}^2a_{32}^2\, ,\qquad
m_2^2+m_3^2=a_{21}^2+a_{31}^2+a_{32}^2\, .
\eg

Without knowledge of the masses $m_2$ and $m_3$ from some other source
there are still various possibilities of choosing $a_{21}$, $a_{31}$, and
$a_{32}$ (keeping in mind, of course, that they are not all
independent). Among these there are two which are particularly interesting.
These are
\begin{itemize}
\item[(A)] $a_{31}=0$, where, say, $m_2^2=a_{21}^2$, $m_3^2=a_{32}^2$,
\item[(B)] $a_{21}=0$, where, say, $m_2=0$, $m_3^2=a_{31}^2+a_{32}^2$.
\end{itemize}

In both cases the matrix $V^{(\nu )}$ is the unit matrix and, by
eq.~(\ref{vmix}), the observable mixing matrix is determined solely
by the {\it charged\/} leptons, $V_{mix}=V^{(l)\,\dagger}$, with
$V^{(l)}$ expressed in terms of $m_i\equiv m_e,m_\mu ,m_\tau $ and of
$d_{ik}$ as worked out in \cite{HS}. Setting
$d_{11}=d_{22}=d_{33}\equiv d$ we find
\bg \label{v}
V_{mix}=\left(\begin{array}{ccc}
 f(m_e)/N_e   & f(m_\mu )/N_\mu & f(m_\tau )/N_\tau \\
 g(m_e)/N_e   & g(m_\mu )/N_\mu & g(m_\tau )/N_\tau \\
 h(m_e)/N_e   & h(m_\mu )/N_\mu & h(m_\tau )/N_\tau
\end{array}\right)\, ,
\eg
where the functions $f$, $g$, and $h$ are given by
\bg \label{f}
f(m_i)=d^2d_{21}d_{32}-dd_{31}e^{-i\Phi^{(l)}}(d^2-m_i^2)\, ,
\eg
\bg \label{g}
g(m_i)=m_i^2d_{21}d_{31}e^{-i\Phi^{(l)}}-dd_{32}(d^2-m_i^2)\, ,
\eg
\bg \label{h}
h(m_i)=(d^2-m_i^2)^2-d_{21}^2m_i^2\, .
\eg
The normalization factors in the denominator are
\bg \label{N}
N_e=\left\{\left[ (d^2-m_e^2)^2-m_e^2d_{21}^2\right]
(m_\tau^2-m_e^2)(m_\mu^2-m_e^2)\right\}^{1/2}\, ,
\eg
$N_\mu$ and $N_\tau$ being obtained from this by cyclic permutation of
$\left\{ m_e,m_\mu ,m_\tau \right\}$. Note that due to the equations
(\ref{lpol1}) - (\ref{lpol3}) there are only two free parameters, say
$d_{21}$ and $d_{32}$.

The form (\ref{v}) of the mixing matrix and the formulae (\ref{f} -
\ref{N}) are already quite restrictive, independently of the choice of 
$d_{21}$ and of $d_{32}$. This is seen as follows. Define the functions
$F_i(z)$, $i=1,2,3$, by
\bg \label{F1}
F_1(z):=\left\{ 
        \frac{(d^2-m_e^2)^2-m_e^2z^2}{(m_\tau^2-m_e^2)(m_\mu^2-m_e^2)}
        \right\}^{1/2}\, ,
\eg
with $F_2$ and $F_3$ obtained from (\ref{F1}) by cyclic permutation of
$\left\{ m_e,m_\mu ,m_\tau \right\}$. Denoting the entries of $V_{mix}$,
eq. (\ref{v}), by $V_{ik}$, we see from eqs.~(\ref{h}) and (\ref{N}) that
the moduli of the elements in the third line are given by
\bg \label{V3M}
\vert V_{3i}\vert = F_i(z=d_{21})\; ,\; i=1,2,3\, .
\eg
Note that because of the ordering of the lepton masses the denominator 
in $F_2$ is negative and, hence, its numerator must also be negative.
{}From these equations one concludes 
\bg \label{d21}
\vert m_\mu^2-d^2\vert /m_\mu\leq\vert d_{21}\vert\leq
(m_\tau^2-d^2)/m_\tau\, .
\eg
When expressed in terms of the $\tau$-mass, this means that
$\vert d_{21}\vert /m_\tau$ must lie between $0.0483$ and $0.9993$. These 
boundary values may in turn be inserted into the moduli to obtain limits
on them, viz.
\bg\label{V3i}
0.01006\leq\vert V_{31}\vert\leq 0.01116,\quad
0\leq \vert V_{32}\vert\leq 0.99995,\quad
0\leq\vert V_{33}\vert\leq 0.99994\, .
\eg
Thus, while $\vert V_{32}\vert$ and $\vert V_{33}\vert$ remain practically
unrestricted, $\vert V_{31}\vert$ must always be very small.

One can also work out limits on the moduli of the elements in the 
{\it first\/} line
of (\ref{v}) which turn out to be even more interesting. In calculating
$\vert f(m_i )\vert^2$ from eq.~(\ref{f}), eliminating $d_{31}$ and
$\cos\Phi$ by means of eqs.~(\ref{lpol2}) and (\ref{lpol3}), and using 
$d^6=m_e^2m_\mu^2m_\tau^2$, one obtains
\bg \label{V1M}
\vert V_{1i}\vert =\left(\frac{d}{m_i}\right)    F_i(z=d_{32})\, ,\quad
(m_1=m_e,m_2=m_\mu ,m_3=m_\tau )\, .
\eg

By the same reasoning as above one sees that $\vert d_{32}\vert$
must also lie in the interval (\ref{d21}). 
The expressions (\ref{V1M}) are essentially the same as (\ref{V3M}), 
except for the
factors $(d/m_e )=89.59$, $(d/m_\mu )=0.4333$, $(d/m_\tau )=0.02576$. 
Because these factors are so different from one another, the allowed
intervals (\ref{V3i}) for the third line, when applied to the {\it first\/} 
line, become
\bg \label{V1i}
0.9013\leq\vert V_{11}\vert\leq 0.9997,\quad
0\leq \vert V_{12}\vert\leq 0.4333,\quad
0\leq\vert V_{13}\vert\leq 0.02576\, .
\eg
Thus, while the narrow interval for the first element is shifted to 
somewhere close
to 1, the very loose constraints on the second and third element of the 
{\it third\/} line are turned into rather tight limits for the corresponding 
elements of the {\it first\/} line. 

These limits as obtained on the assumption that neutrinos have no primordial 
masses and $V^{(\nu)}=\id$ (so that $V_{mix}$ stems from the charged 
leptons only), could already be tested by experiment, either confirming
compatibility with this simple scheme or disproving it. Clearly, a more
detailed analysis will have to include a thorough numerical analysis of 
acceptable values for $d_{21}$ and $d_{32}$\footnote{We note in passing that
$d_{21}$ and $d_{32}$ in fact are not completely independent. Using unitarity
of $V_{mix}$, one concludes from the moduli of the elements in the second
line that the combination $d_{21}^2+(d^2/m_\mu^2)d_{32}^2$ is bounded from
above by $0.9992m_\tau^2$, while the combination
$d_{21}^2+(d^2/m_\tau^2)d_{32}^2$ is bounded from below by $0.00287m_\tau^2$.
These bounds are so mild that they will not play a decisive role.} 
and/or other choices of the
neutrino mass matrix.
\\[4pt]
{\bf 4.} As an illustration of this general framework we briefly describe
the mass matrices as obtained in the framework of the Mainz-Marseille
model (cf. introduction). In this model 
the left-chiral doublet and the right-chiral singlet of a given lepton
family fit into the fundamental representation
\bg  \label{frep}
[Y_0=-1,I_0=\frac{1}{2}]\longrightarrow
(I=\frac{1}{2},Y=-1)\oplus (I=0,Y=-2)\, ,
\eg
where the right-hand side gives the decomposition in terms of
$Lie\left( SU(2)\times U(1)\right)$. A right-handed neutrino would be
classified according to the trivial representation $[Y_0=0,I_0=0]
\equiv (I=0,Y=0)$. As discussed in \cite{CES} the two representations can
be combined to a reducible but indecomposable representation
\bg \label{lrep}
\rho_l:=[Y_0=-1,I_0=\frac{1}{2}]\lplus [Y_0=0,I_0=0] \, ,
\eg
which describes a charged Dirac lepton $l$ and its neutrino $\nu_L^{(l)}$,
along with a right neutrino $\nu_R^{(l)}$ which does not couple directly
to the gauge bosons but does couple to the leptonic multiplet
$(l_L,l_R,\nu_L)$ via the Higgs doublet. The next simplest, and physically
interesting possibility is to classify two or three lepton families in
the reducible indecomposable representations which are obtained by taking
semi-direct sums of the representation $\rho_l$, eq.~(\ref{lrep}),
according to the pattern shown in eq.~(\ref{semi}).
While (\ref{lrep}), in a natural way, allows to introduce a (Dirac) mass of the
neutrino $\nu_l$, without introducing right-handed interactions, the
representation (\ref{semi}) provides in addition family mixing both in the
charged and in the neutral lepton sector. Indeed, in a representation with
three families the even and odd generators of $su(2\,\vert\, 1)$ have
the form, respectively,
\begin{eqnarray} 
\widehat{T}_i & = & \left(\begin{array}{ccc}
(T_i)^{11} & 0 & 0\\(T_i)^{21} & (T_i)^{22} & 0 \\  \label{bloc}
(T_i)^{31} & (T_i)^{32} & (T_i)^{33}\end{array}\right)\nonumber\\
\widehat{\Omega}_i & = & \left(\begin{array}{ccc}
(\Omega_i)^{11} & 0 & 0\\(\Omega_i)^{21} & (\Omega_i)^{22} & 0\\
(\Omega_i)^{31} & (\Omega_i)^{32} & (\Omega_i)^{33}\end{array}\right)\, .
\end{eqnarray}

The entries of these block triangular matrices are $4\times 4$-matrices. In
the basis
\bg \label{psi}
\Psi =\left(\psi^{(1)},\psi^{(2)},\psi^{(3)}\right)^T\, ,\mbox{with }
\psi^{(i)}=\left( l_L^{(i)},\nu_L^{(i)},l_R^{(i)},\nu_R^{(i)}\right)^T\, ,
\eg
and using Pauli matrices $\tau_i$ and projection
operators $P_\pm =(1\pm\tau_3)/2$, the odd generators must have the form
\bg \label{opl}
(\Omega_+)^{ik}=\left(\begin{array}{cc}
\mr{0} & a_{ik}P_-\\ \alpha_{ik}P_+ & \mr{0}\end{array}\right)\quad
(\Omega_-^\prime)^{ik}=\left(\begin{array}{cc}
\mr{0} & d_{ik}P_+\\ \delta_{ik}P_- & \mr{0}\end{array}\right)\,
\eg
\bg  \label{omi}
(\Omega_-)^{ik}=\left(\begin{array}{cc}
\mr{0} & b_{ik}\tau_+\\ \beta_{ik}\tau_+ & \mr{0}\end{array}\right)\quad
(\Omega_+^\prime)^{ik}=\left(\begin{array}{cc}
\mr{0} & c_{ik}\tau_-\\ \gamma_{ik}\tau_- & \mr{0}\end{array}\right)\, ,
\eg
the bold face $\mr{0}$ denoting the $2\times 2$ zero matrix.
As we showed in \cite{HS} it is no restriction of generality to choose
the even generators block diagonal, i.e.
\bg \label{IY}
\left( I_k\right)^{ik}=0=\left( Y\right)^{ik}\, ,\;\mbox{for }
(ik)=(21),(31),(32)\, . \eg

In the basis chosen the diagonal blocks are
\[ (I_3)^{ii}=\diag (-\frac{1}{2},\frac{1}{2},0,0)\, ,\;
   \sqrt{2}(I_\pm )^{ii} =\left(\begin{array}{cc}
   \tau_\mp & \mr{0}\\ \mr{0} & \mr{0} \end{array}\right)\, ,\;
   (Y)^{ii}=\diag (-1,-1,-2,0)\, .
\]

The interaction of the lepton state (\ref{psi}) with the gauge bosons of
$SU(2)\times U(1)$ and the Higgs field is encoded in the superconnection of
the model,
\bg \label{con}
{\cal A}=i\left\{ a\widehat{\mb{I}}\cdot\mb{W}+
\frac{1}{2}b\widehat{Y}W^{(8)}+\frac{1}{\mu}
\left[ \widehat{\Omega}_-^\prime\phi^{(0)}+\widehat{\Omega}_+^\prime\phi^{(+)}
+\widehat{\Omega}_+\overline{\phi^{(0)}}+\widehat{\Omega}_-\overline{\phi^{(+)}}
\right]\right\}\, ,
\eg
with $a,b$ real, dimensionless parameters, $\mu$ the mass scale of
the theory. In particular, and as explained in ref.~\cite{HS},
the mass terms which follow
from eq.~(\ref{con}) are
\bg \label{mass1}
\LD_{mass}  =  \frac{1}{2}\left[
 \left(\overline{\Psi}\widehat{\Omega}_+\Psi\right)_{\nu}+
 \left(\overline{\Psi}\widehat{\Omega}_-^\prime\Psi\right)_{l}
 \right] +\mbox{h.c.}\, ,
\eg
where the indices indicate that neutrinos only get contributions from
$\widehat{\Omega}_+$, the charged partners get contributions only from
$\widehat{\Omega}_-^\prime$. Thus, in this model $\LD_{mass}$ has indeed 
the general form
of eqs.~(\ref{mass2}) and (\ref{tria}) above.

The graded commutators of the even and odd generators of the algebra
yield relations between the complex parameters
$a_{ik}\ldots\gamma_{ik}$. These read\footnote{Note that the analogs of
eqs.~(\ref{par34}) with $\delta_{ik}$ replaced by $\alpha_{ik}$,
and $a_{mn}$ by $d_{mn}$, exist but, by (\ref{par32}), are linearly
dependent on eqs.~(\ref{par35}).}
\bg \label{par30}
b_{ik}=a_{ik},\; \beta_{ik}=-\alpha_{ik},\;
c_{ik}=-d_{ik},\; \gamma_{ik}=\delta_{ik}\, ,\quad i,k=1,2,3\, ,
\eg
\bg \label{par31}
a_{ii}\delta_{ii}=0\, ,\quad i,k=1,2,3\, ,
\eg
\bg\label{par32}
d_{ii}\alpha_{ii}=-1\, ,\quad i,k=1,2,3\, ,
\eg
\bg \label{par33}
   a_{21}\delta_{11}+a_{22}\delta_{21}=0 \, , \quad
   a_{32}\delta_{22}+a_{33}\delta_{32}=0 \, , \quad
   \sum_{i=1}^3 a_{3i}\delta_{i1}=0\, ,
\eg
\bg \label{par34}
   \delta_{21}a_{11}+\delta_{22}a_{21}=0 \, , \quad
   \delta_{32}a_{22}+\delta_{33}a_{32}=0 \, , \quad
   \sum_{i=1}^3 \delta_{3i}a_{i1}=0\, ,
\eg
\bg \label{par35}
   d_{21}\alpha_{11}+d_{22}\alpha_{21}=0 \, ,\quad
   d_{32}\alpha_{22}+d_{33}\alpha_{32}=0 \, ,\quad
   \sum_{i=1}^3 d_{3i}\alpha_{i1}=0\, .
\eg

We note, in particular, that the right-hand sides of 
eqs.~(\ref{par31}) and (\ref{par32})
are the electric charges of the neutrinos and their charged partners,
respectively. The former are important in discussing the remaining
freedom in choosing the parameters $a_{ik}$ and give an interesting
interpretation of the cases (A) and (B) discussed above. Indeed,
the assumption $a_{ii}=0$ complies with the condition (\ref{par31})
which says that either $a_{ii}$ or $\delta_{ii}$ have to vanish because
neutrinos have no electric charge. Eqs.~(\ref{par33}) and (\ref{par34})
then reduce to, respectively,
\bg \label{par301}
   a_{21}\delta_{11}=0 \, , \quad
   a_{32}\delta_{22}=0 \, , \quad
   a_{31}\delta_{11}+a_{32}\delta_{21}=0\, ,
\eg
\bg \label{par302}
   \delta_{22}a_{21}=0 \, , \quad
   \delta_{33}a_{32}=0 \, , \quad
   \delta_{32}a_{21}+\delta_{33}a_{31}=0\, .
\eg

Barring the trivial case where all $a_{ik}$ are zero, eqs.~(\ref{par301} - 
\ref{par302}) admit the following choices\\
(i) $a_{21}\not= 0$, $a_{32}\not= 0$: In this case all $\delta_{ij}$ except
$\delta_{31}$ must vanish. Two neutrinos are massive, one remains massless.
Depending on whether $a_{31}=0$ the neutrino mass matrix is diagonal and
the mixing matrix (\ref{vmix}) is determined by the mass matrix of charged
leptons only.\\
(ii) $a_{21}\not= 0$, $a_{32}=0$: In this case $\delta_{11}$ and
$\delta_{22}$ must vanish, $\delta_{21}$ and $\delta_{31}$ remain undetermined,
while $\delta_{32}$ and $\delta_{33}$ are related by
\[
\delta_{32}=-\frac{a_{31}}{a_{21}}\delta_{33}\, .
\]
Two neutrinos are massive and mix, as long as $a_{31}\not= 0$. If $a_{31}=0$
one more neutrino becomes massless and, obviously, there is no mixing.\\
(iii) $a_{21}=0$, $a_{32}\not= 0$: Here $\delta_{22}=\delta_{33}=0$,
\[
\delta_{21}=-\frac{a_{31}}{a_{32}}\delta_{11}\, ,
\]
while $\delta_{32}$ and $\delta_{31}$ remain undetermined. Two neutrinos are
massless, the mass matrix is diagonal. The mixing matrix (\ref{vmix})
is determined by the charged sector.\\
(iv) $a_{21}=a_{32}=0$, but $a_{31}\not= 0$: Here $\delta_{11}=\delta_{33}=0$,
all others remain undetermined. Two neutrinos are massless, the mass matrix
is diagonal.

Note that cases (ii) -- (iv) are not simply limiting cases of case (i) because
in the latter all $\delta_{ik}$ but $\delta_{31}$ must vanish whereas in
cases (ii) to (iv) some of these parameters may be different from zero.
As we noted in sec. 3 the physically interesting cases are (A) and (B) which
are contained in (i) and (iii).\\[4pt]
{\bf 5.} As CC weak interactions involve left-chiral fields only, triangular 
mass matrices of the type (\ref{tria}) for neutrinos
and their charged partners provide the most general, yet concise, description
of leptonic state mixing. It seems plausible that in the absence of
electroweak interactions, leptons are mass-degenerate in each charge sector
and that neutrinos possess no primordial masses. 
These assumptions are equivalent to the choice (\ref{degen}) for the
diagonal entries of $T^{(\nu )}$ and $T^{(l)}$. The characteristic
polynomial for the mass matrix $\left( T^{(\nu )}T^{(\nu )\,\dagger}\right)$
allows to classify the possible choices for the remaining parameters
$a_{21}$, $a_{31}$, and $a_{32}$. Among these the choices (A) and (B) are
particularly interesting because in these cases the neutrino mass matrix
is diagonal and, therefore, the mixing matrix is determined by the mass 
matrix of {\it charged\/} leptons only. If the atmospheric neutrino
experiment or the short-baseline oscillation experiment is
confirmed, then case (B) is excluded because one needs (at least) two
nonvanishing mass differences. Note that even though, by our assumption of
vanishing primordial neutrino masses (\ref{degen}), one neutrino stays 
massless there is nevertheless mixing between all three families. This is
due to the fact that the electron has a finite mass and, in the charged 
lepton's mass matrix, does not decouple from the other two families.

The example of two families where all parameters are fixed by the masses, 
shows that mixing is compatible with the specific resonance
oscillation needed to explain the solar neutrino fluxes, cf. 
eq.~(\ref{sin2}). With three families there is more, but not much more, 
flexibility. In the case where the mass matrix of neutrinos is such that
$V^{(\nu )}=\id$, i.e. where $V_{mix}$ depends on the charged 
lepton sector only, there are two free parameters bounded by the
interval (\ref{d21}) (and some mild compatibility condition among them). 
However, the scheme is so tight that even without fixing these parameters,
the moduli of the elements in the third and first lines of $V_{mix}$
must lie in the intervals (\ref{V3i}) and (\ref{V1i}), respectively.

In a more general situation where $V^{(\nu )}$ is not proportional to the
unit matrix $\id$ the mixing matrix (\ref{vmix}) will also depend on the
neutrino masses. We hope to return to a more detailed numerical analysis
in a future publication.

Finally, we illustrate this general
analysis by a model that is an extension of the standard model within
noncommutative geometry. This model sheds new light on the classification
of parameter choices by relating them to graded commutators of generators
of the underlying graded Lie algebra, and, thereby, establishes a link
to lepton quantum numbers.

\vspace*{2cm}

F. Scheck is grateful for the kind hospitality extended to him by 
Harold Fearing and the theory group 
at TRIUMF, Vancouver, where part of this work was done.


\end{document}